\documentclass[preprint,showpacs,preprintnumbers,amsmath,amssymb]{revtex4}

\usepackage{graphicx}
\usepackage{dcolumn}
\usepackage{bm}
\newcommand{\m}{\mathbf}
\newcommand{\h}{\hat}
\newcommand{\n}{\nonumber}

\begin{document}
\title{Surface  monolayers and   magnetic field}

\author{S. V. Maleyev}
\affiliation{National Research Center
Kurchatov Institute Petersburg Nuclear Physics Institute, Gatchina, St.\ Petersburg 188300, Russia}
 \large

\begin{abstract}
 We study theoretically the  magnetic properties of the surface monolayers with the antiferromagnetic  (AF) and ferromagnetic (FM) exchange interactions where the Dzyaloshinskii-Moriya interaction (DMI) is a result  of the mirror  symmetry breaking.  To study the DMI helices in magnetic field  a  method is proposed. In zero field the  DMI  gives rise  a cycloid in both AF and FM cases. The cycloid orientation is determined by the DMI induced  in-plane  anisotropy   with the symmetry of the layer lattice. As a result
we have one, two and three chiral domains in the rectangular, square and triangular lattices respectively. The magnetic structure  of the  $M n/W(1 1 0)$ monolayer is explained.
The out-of-plane anisotropy may restore a collinear magnetic order.
The chiral domains are rotated by the in-plane field.   In some field directions  the spin flops are predicted.
In the out-of-plane	field the chirality  follows the field direction. The length of the   cycloid wave-vector decreases.
In the perpendicular field there is the spin flop to the corresponding collinear state.
A possibility of the layer  electric polarization is discussed.          	
\end{abstract}

\maketitle

\section{Introduction}

Magnets  with the Dzyaloshinskii-Moriya interaction (DMI)
 \cite{D,MT}  have many  features  unknown  in the conventional magnetic systems. Some of them remain unexplained.
  We mention  A-phase and Skyrmion lattice in B20 magnets \cite{LB,PF,G2} and  the electric polarization flops  in the multiferroics (See\cite{MU} and references therein).

The ultra thin magnetic films and interfaces represent a special  class of the DM  magnets where the DMI  is a result of the mirror symmetry breaking \cite{L}.

 We are interested in the single surface layers. The principal experimental results are following:

 $M n/W(1 1 0)$ layer is the antiferromagnetic   (AF) cycloid \cite{BO,SE}.
  $Fe/W(1 1 0)$ layer is a ferromagnetic (FM) with  the spins in the surface  \cite{EL,W}.  $Fe/W(0 0 1)$ and $Fe/Ir(0 0 1)$ layers. Both are antiferromagnetics   \cite{KU,KUD}. In the first case  the spins  are perpendicular to the surface.

In this paper we study  theoretically   magnetic properties of the surface monolayers with the antiferromagnetic  (AF) and ferromagnetic (FM) exchange interactions where the DMI is a result of the mirror  symmetry breaking \cite{L}. The principal results are following.

 To study magnetic field behavior of  the DMI helices a  method is proposed.

In zero field the  DMI  gives rise  a cycloid in both AF and FM cases. The cycloid orientation is determined by the DMI induced  in-plane  anisotropy   with the symmetry of the layer lattice. As a result
we have one, two and three chiral domains in the rectangular, square and triangular lattices respectively. The magnetic structure  of the  $M n/W(1 1 0)$ monolayer is explained.

The out-of-plane anisotropy may restore a collinear magnetic order.

The chiral domains are rotated by the in-plane field.   In some field directions  the spin flops are predicted.

In the out-of-plane	field the chirality  follows the field direction. The length of the   cycloid wave-vector decreases.
In the perpendicular field there is the spin flop to the corresponding collinear state. The spin flops  in frustrated helices were considered in \cite{U}.
	
 A possibility of the layer  electric polarization is discussed. It may  appear in the cycloidal state as in multiferroics \cite{MU,SD}.

The paper is organized as follows.  In Sec.II the  model is described.  General expressions are derived for the energy of the DMI helices  in the magnetic field. In Sec.III the rectangular  AF  and FM layers are studied.  The uniaxial anisotropy is considered in Sec.IV.  Sec.V and VI are devoted to the square and triangular lattices respectively. A possibility of the layer electric polarization is considered in Sec.VII. In the last Sec.VIII we discus a role  of the DMI in the films with few layers.
     	
\section{Model}

We derive below general expressions for the classical energy of the helices with the DMI in the  magnetic field.
 Corresponding Hamiltonian is following
\begin{equation}
H=(1/2)\sum\{J_\m{R,R'}(\m{S_R\cdot S_{R'}})+(\m D_{\m R{\m R'}}\cdot[\m S_{\m R}\times\m S_{\m {R'}}])\}+\sum(\m{H\cdot S_R}),
\end{equation}
where  $\m D_{{\m R'}\m R}=-\m D_{\m R{\m R'}}$. The last term is the Zeeman energy.

In the surface monolayer the DMI is a result of the mirror symmetry breaking.  In this case  the DMI  must  be on each bond $\m b$ connecting two spins \cite{MT,L}. Neglecting the substrate structure we have \cite{L}
\begin{equation}
\m{D_b}=d_\m b[\h z\times\m b],\;\m D_{-\m b}=-\m{D_b}.
\end{equation}
where $\h z$ is the unit vector perpendicular to the surface.

The DMI  distorts the commensurate magnetic order and a helical structure may appear.
To describe it we use the classical part of the Kaplan representation \cite{K}
\begin{equation}
 \m{S_R}= S( \m  A e^{i\m{k\cdot R}}+{\m A}^* e^{-i\m{k\cdot R}})\cos\alpha+S\h c\sin\alpha,
\end{equation}
where $\m A=(\h a-i\h b)/2$, unit vectors   $\h a\perp\h  b$, $[\h a\times\h b]=\h c$  and $\alpha$ is the cone angle. We have
\begin{equation}
	\m{(A\cdot A)}=0,\;\m{(A\cdot A^*)}=1/2,\;
[\h c\times\m A]=i\m A,\; \m{[A\times A^*]}=i\h c/2.
 \end{equation}
These expressions contain six free parameters: wave-vector $\m k$, unit vector $\h c$ and cone angle $\alpha$.   If $\alpha=0$ at $\h c ||\m k$ and $\h c\perp\m k$ we have the planar helix and cycloid respectively.

The vectors $\m M=S\h c \sin\alpha$  and $\m C=S^2\h c\cos^2\alpha$ are the helix   magnetization and the chirality respectively. They have different $t$-parity as the   spin is $t$-odd.

		From Eqs.(1-4) we obtain the classical energy of the helix
\begin{align}
E= (S^2/2)\{J_0\sin^2\alpha+[J_\m k+i(\m{D_k}\cdot\h c)]\}\cos^2\alpha+S(\m H\cdot\h c)\sin\alpha,\\
J_\m k=\sum_\m b J_\m b\cos\m{k\cdot b},\;\m D_\m k=i\sum_\m b d_\m b [\h z\times\m b]\sin\m{k\cdot b}.
\end{align}
The  helical spin structure is  determined by  minimum of the energy (5). From $d E/d\alpha=0$ we obtain
\begin{align}
\sin\alpha&=-\frac{(\h c\cdot\m H)}{H_c},\;H_c=S J_0-S[J_\m k+i(\h c\cdot\m{D_k})],\\
E&=(S^2/2)[J_\m k+i(\h c\cdot\m{D_k})]-\frac{S(\h c\cdot\m H)^2}{2H_c}.
\end{align}
We consider below the antiferromagnetic (AF) and ferromagnetic (FM)  exchange interactions. In the first case  one must replace $\m k\to\m k_{A F}+\m k$ where $\m k_{AF}=(\pi,\pi)$ is the AF part of the wave-vector. In the second case we must replace $J\to-J$. As a result we obtain
\begin{align}
E_a&=-(S^2/2)[J_\m k+i(\h c\cdot\m{D_k})]-\frac{S(\h c\cdot\m H)^2}{2H_a} \; (AF),\\
E_f&=-(S^2/2)[J_\m k-i(\h c\cdot\m{D_k})]-\frac{S(\h c\cdot\m H)^2}{2H_f}\;  (FM),
\end{align}
where
\begin{align}
H_a&=S[J_\m k+i(\h c\cdot\m{D_k}+J_0)]\simeq 2S J_0;\n\\
H_f&=S[J_\m k-i(\h c\cdot\m{D_k})-J_0]\sim S J_0 k^2,
\end{align}
where in r.h.s. we put $k\ll 1$. Expressions for  $H_f$  will be given below.

Eqs. (9-11) do not change if  one replace $(\h c,\m k)\to (-\h c,-\m k)$. So the energy has two minima, but we consider them below as one state.

We note that Eqs.(5) and (9-11) do not depend on the special form of the DMI.

   General expressions for $\m H$ and $\h c$ used below
 are following
\begin{align}
\m H&=H(\cos\Psi\sin\Theta,\sin\Psi\sin\Theta,\cos\Theta),\n\\
  \h c&=(\cos\phi\sin\theta,\sin\phi\sin\theta,\cos\theta).
\end{align}

\section{Rectangular lattice}

We consider the rectangular layer with the nearest neighbor  (n.n.) interaction shown in FIG.1. It is a model  for $M n/W(1 1 0)$ and  $F e/W(1 1 0)$ layers studied  in \cite{BO,SE,EL,W}. In the $M n$ case the DMI  can explain   observed spin structure but in the $Fe$ case one must add the uniaxial anisotropy (See Sec.IV).

		\begin{figure}
			 \centering
				 \includegraphics[width=14cm]{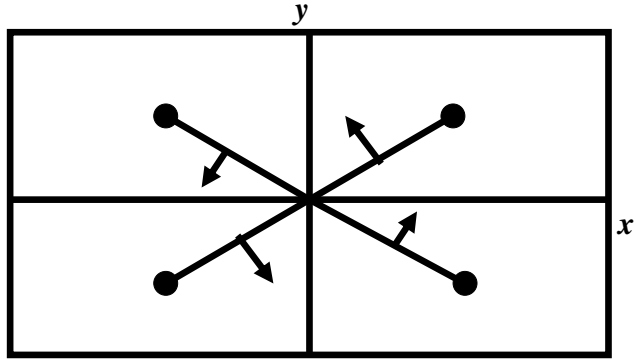}
			 \caption{
	The nearest neighbors in the $M n$ ane $F e$ monolayers on the $W(1 1 0)$ plane.  The lines and arrows are the n.n. bonds and DM vectors.
The lattice constants along $x$ and  $y$ axes  are $\sqrt{2}$
			and unity respectively. }
			 \label{Fig.1}
		 \end{figure}

\subsection{AF layer}

According FIG.1 for   four   n.n. bonds  we have $\m b=\pm(\h x\sqrt{2}\pm\h y)/2)$. In zero field $\h c$ is in the  $(x,y)$ plane and $\theta=\pi/2$ in Eq(12).
 If $\m k=k(\cos\xi,\sin\xi)$ from Eqs.(6) and (9) we obtain
\begin{align}
  E_a=- S^2 J\{\cos k \cos(\xi-\rho)+\cos k\cos(\xi+\rho)\n\\
	-k_0[\sin(\phi-\rho)\sin k\cos(\xi-\rho)+
	\sin(\phi+\rho)\sin k\cos(\xi+\rho)]\},
	\end{align}
	where $k_0=d/J$,  $\cos\rho=\sqrt{2/3}$  and $\sin\rho=1/\sqrt{3}$ \cite{CC}.   In the $k^2$ approximation   we obtain
	\begin{align}
E_a=-(S^2 J_0/2)\{1-k^2(\cos^2\rho \cos^2\xi+\sin^2\rho \sin^2\xi)/2-\n\\
k k_0[\cos^2\rho\sin(\phi-\xi)+ \cos 2\rho\cos\phi\sin\xi ]\}
	\end{align}
	The minimum conditions are following
	\begin{align}
	k(\cos^2\rho\cos^2\xi+\sin^2\rho\sin^2\xi)+k_0[\cos^2\rho\sin(\phi-\xi)+\cos 2\rho\cos\phi\sin\xi ]&=0,\n\\
	-k\cos 2\rho\sin\xi\cos\xi-k_0[\cos^2\rho\cos(\phi-\xi)-\cos 2\rho\cos \phi\cos\xi]&=0.
	\end{align}
These equations have  two	solutions: $k=k_0,\xi=\phi+ \pi/2$ and $k=-k_0,\xi=\phi-\pi/2$. Both give the same  result .  As $\cos^2\rho=2/3$ we obtain
\begin{align}
 E_a&=-(S^2 J_0/2)\left[1+\frac{k^2}{6}(1+\sin^2\phi)\right]-\frac{S H^2}{2H_a}\cos^2(\phi-\Psi), \\
\m k&=k_0(-\sin\phi,\cos\phi),\;(\h c\cdot\m k)=0,\;[\h c\times\m k]=k_0\h z,
\end{align}
where in Eq.(16) $J_0=4J$ and    the energy of the in-plane field is add [See Eqs.(9) and (11)].  In Eq.(16) the first term is the classical energy of the AF state.  It may be omitted.  The DMI is represented  by  next two terms where
the $\sin^2\phi$ term is the DMI induced in-plane
anisotropy.

In zero field  $\sin^2\phi=1$, $\m k=k_0(-1,0) ||\h x$ and $\h c=(0, 1)||\h y$
  \cite{A}. It is the AF  cycloid  observed in \cite{BO,SE} as  $x$ axis in FIG.1 is  the  $(1 1 0)$ direction in the $W$ plane.


The in-plane field rotates the $(\h c,\m k)$ cross.
If $\Psi=0\; (\m H||\h x)$ from Eq. (16) we obtain
\begin{equation}
E_{a 1}=-(S^2 J_0 k^2/12) \{1+[1-(H/H_{a c})^2]\sin^2\phi+(H/H_{a c})^2\},
\end{equation}
where we omitted $-S^2 J_0/2$ term and  $H_{a c}=H_a k/2\sqrt{6}$. If $H<H_{ac}$ this energy  is minimal at $\sin^2\phi=1$ an $\sin\alpha=0$ [See Eq.(7)].  At   $H=H_{a c}$ there is the first order  spin flop transition to the conical cycloid with $\sin^2\phi=0,\;\h c=(1,0),\; \m k=k_0(0,1)$ and $\sin\alpha=-H/H_c$ [See FIG.2a,b].

\begin{figure}
	\centering
		\includegraphics[width=18cm]{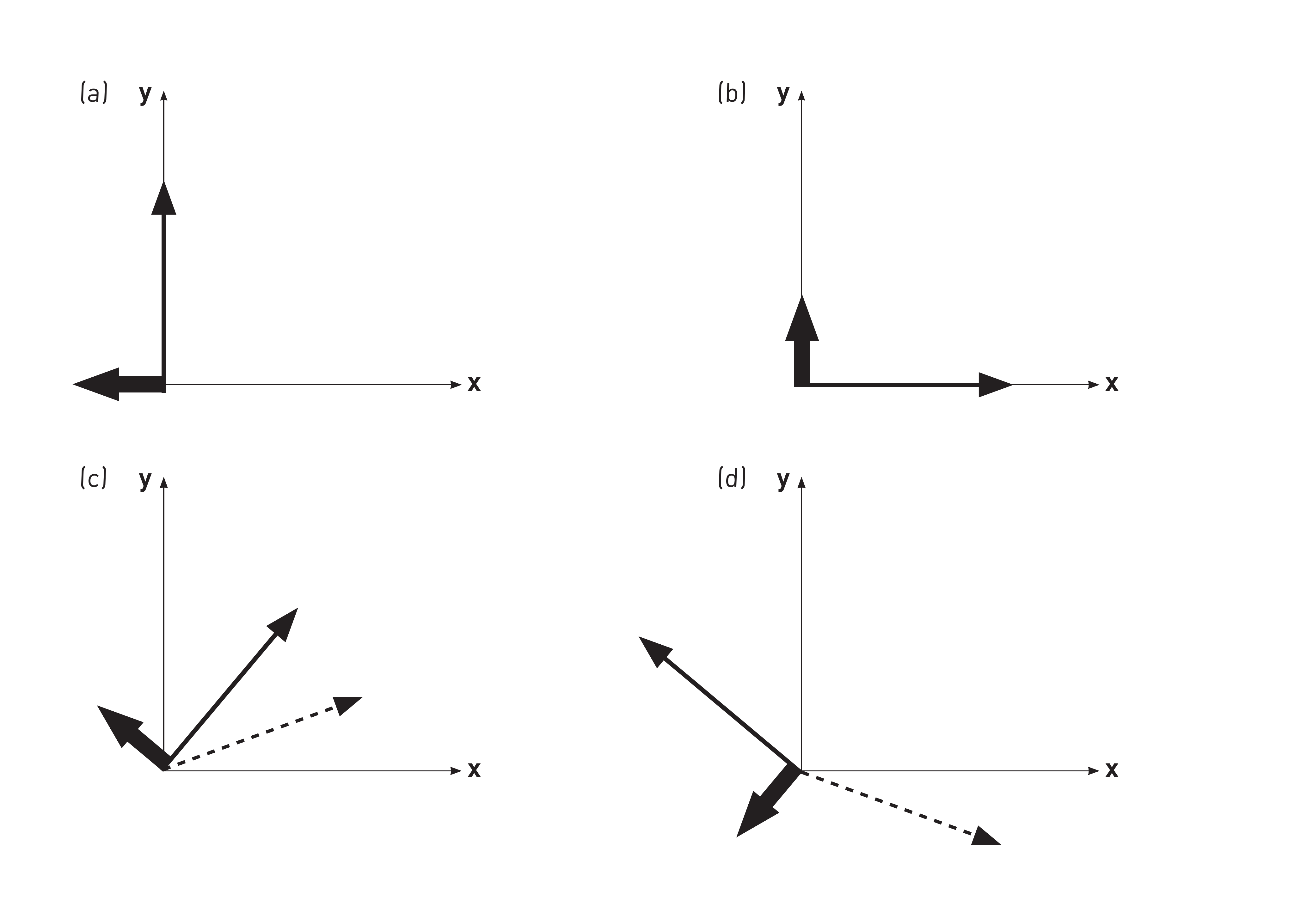}
	\caption{The field dependence of the $(\h c,\m k)$ orientation. (a) $\Psi=0,\; H<H_{a c}$. (b) $\Psi=0,\; H>H_{a c}$.  (c) $\Psi>0$. (d) $\Psi<0$.  Thin, thick and dashed arrows are $\h c,\;\m k$ and $\m H$ respectively.}
	\label{Fig.2}
\end{figure}

	 If $\Psi=\pi/2,\;(\m H  || \h y)$ the energy is minimal at $\phi=\pi/2$. We have the conical cycloid with  $\sin\alpha=-H/H_a$. In both cases there  is the spin flip at $H=H_a$.

In general case  the minimum conditions of the energy  (16) are the same as in Appendix A  if one replaces $(\theta,\Theta)\to(\phi,\Psi)$.
From  Eqs.(A6), (A9)  and (A10) we obtain
\begin{align}
E_{a 1}&=-(S^2J_0 k^2/24
)[3+g+(1-2g\cos 2\Psi+g^2)^{1/2}],\n\\
\sin^2\phi&=\frac{1}{2}\left[1+\frac{1-g\cos 2\Psi}{(1-2g\cos 2\Psi+g^2)^{1/2}}\right],\n\\
\sin 2\Psi/\sin 2\phi&>0, 	
\end{align}
where $g=(H/H_{a c})^2$. The last inequality determines the $\phi$ dependence on sign of $\Psi$ as shown in FIG.2(c,d). At $\Psi=0$ these expressions describe the  spin flop as $(1-2g\cos 2\Psi+g^2)^{1/2}\to|1-g|$. If $g\gg 1$ we have $\sin^2\phi\simeq\sin^2\Psi$.

In the out-of-plane field the in-plane part of $\h c$  has the factor $\sin\theta$ [See Eq.(12)]. As a result  in Eq.(13) one must replace $k_0 \to k_0\sin\theta$  and we obtain
\begin{align}
E_{a 1}&=-\frac{S^2J_0 k^2}{12}\left\{(1+\sin^2\phi)\sin^2\theta+\frac{H^2}{H^2_{a c}}[\cos(\phi-\Psi)\sin\theta \sin\Theta+\cos\theta \cos\Theta]^2\right\},\\
\m k&=k_0(-\sin\phi,\cos\phi)\sin\theta.
\end{align}

In the perpendicular field ($\m H||\h z ,\Theta=0$) and $\sin^2\phi=1$
 we have
\begin{equation}
E_{a 1}=-(S^2 J_0 k^2/6)[(1-H^2/2H^2_{a c})\sin^2\theta+H^2/2H^2_{a c}].
\end{equation}
   This  equation describes the first order spin flop transition at $H=H_{s f}=H_{a c}\sqrt{2}$
	from the cycloid to the AF state with the  spins in the  ($x,y$) plain.

In general case  ($\Theta\neq 0$) $\theta$ and $\phi$ are complicated functions of $\Theta,\;\Psi$ and $H$.
In Appendix B is show that in the strong field ($H\gg H_{ac}$) $\theta\simeq\Theta$ and $\phi\simeq\Psi$.  As a result we have the conical cycloid with $\h c||\m H,\; \sin\alpha=-H/H_a$,  $\m k= k_0(-\sin\Psi,\cos\Psi)\sin\Theta$ and $|\m k|=|k_0|\sin\Theta<|k_0|$.  So in the out-of-plane field  the length of $\m k$ decreases.

\subsection{FM  layer}

From Eqs.(7,10) and (11)  in the in-plane field    we have
\begin{align}
E_f&=-\frac{S^2 J_0}{2}\left[1+\frac{k^2}{6}(1+\sin^2\phi)\right]-\frac{S H^2\cos^2(\phi-\Psi)}{2H_F(1+\sin^2\phi)},\\
  H_F&=S J_0 k^2/6,\;\m k=k_0(\sin\phi,-\cos\phi), \\
\sin\alpha&=-\frac{H\cos(\phi-\Psi)}{H_F(1+\sin^2\phi)},
\end{align}
where $k_0=d/J$.  In zero field  $\sin^2\phi=1$.  We have the  FM cycloid with $\h c ||\h y$. However the wave vector $\m k$ has  other sign than in the AF case [See Eq.(17)].


At $\m H\neq 0$ we consider two case:$\Psi=0\;(\m H ||\h x)$ and $\Psi=\pi/2\;(\m H ||\h y)$.  In both cases  we must compare the energy (23) with the energy $E_{F M}=-
S^2 J_0/2-S H$ of the ferromagnetic with the spins along the field. The restriction $\sin\alpha>-1$ must be taken into account too.

  i.$\Psi=0 \;(\m H ||\h x)$.  Equation  $dE_a/d\phi=0$ has three solutions:   $\sin^2\phi=1,\;\sin\phi=0$ and $1+\sin^2\phi=H\sqrt{2}/H_F$. The energy is minimal at  $\sin^2\phi=1$.  We have $E_{f 1}=-S H_F$ and $\sin\alpha=0$.     The first order FM spin flop takes place at $H=H_F$.

ii. $\Psi=\pi/2\;(\m H ||\h y)$.  The energy is minimal at $\sin^2\phi=1$. We have the conical cycloid with $E_{f 1}=-S H_F[1+(H/2H_F)^2]$ and $\sin\alpha=-H/2H_F$.

 In  the perpendicular field  ($\Theta=0$) we have
\begin{align}
E_{f 1}&=-\frac{S H_F}{2}
 \left[(1+\sin^2\phi)\sin^2\theta+\frac{H^2\cos^2\theta}{H^2_F(1+\sin^2\phi)\sin^2\theta}\right],\\
\sin\alpha&=-\frac{H\cos\theta}{H_F(1+\sin^2\phi)\sin^2\theta}.
\end{align}

This  energy has  two extrema.  1) $\sin^2\theta=1,\; E_1=-S H_F(1+\sin^2\phi)/2$ and $\sin\alpha=0$. 2) $\sin^2\theta=H/H_F(1+\sin^2\phi)$ and $E_2=E_1(1-\cos^4\theta)$.
So we have $\sin^2\phi=1,\;E_{f 1}=-S H_F$ and the FM spin flop at $H=H_F$.

\section{Uniaxial anisotropy}

In the rectangular lattice the DMI gives rise a cycloid. The same takes  place  in the square  and triangular lattices considered below.
The AF cycloid  was observed in \cite{BO,SE}. In other cases the FM and AF magnetic structures were found (See Sec.I).

 We demonstrate now that the anisotropy in $\h z$   direction can restore a collinear magnetic order.

 The uniaxial anisotropy is determined as follows
	\begin{equation}
	 H_A=A\sum (S^z_\m R )^2,
	\end{equation}
	where  $A>0$ and $A<0$ correspond to the easy plane and easy  axis anisotropy respectively.	Using Eqs.(3) and (4)  in zero field ($\alpha=0$)    we obtain
\begin{equation}
E_A=(S^2 A/2)(\h a^2_z+\h b^2_z)=(S^2 A/2)\sin^2\theta,
\end{equation}
as $\h a_z^2+\h b_z^2+\h c_z^2=1$ and $\h c_z=\cos\theta$.	

This energy must be added to the cycloid energy $E_1$. According Eqs.(23) and (26) $E_{a 1}=E_{f 1}=-(S^2J_0/2) Q k^2\sin^2\theta$, where $Q=1/3$.  The same expressions take place in the square and triangular lattices with $Q=1/4$ (See Sec. V and VI).
 For the sum $E_1+E_A$ we obtain
\begin{equation}
E=S^2(- J_0 Q k^2/2+A/2)\sin^2\theta.
\end{equation}
We have a cycloid if this energy is lesser than the anisotropic energy   $<E_A>$ in the  collinear state. If $<E_A> <E$ we   have the AF or FM state depending on a type of the exchange interaction.

We have $<E_A>=0$ and $<E_A>=-S^2 |A|$  in the easy plane and easy axis cases respectively.  In both cases the cycloid is stable if
\begin{equation}
J_0 Q k^2>|A|.
\end{equation}
In this expression both sides are  of the second order  of the spin orbit interaction as the DMI  is of the first order
  \cite{MT,L}.

\section{Square lattice}

The nearest neighbor bonds and the DM vectors are shown in FIG.3. We obtain    	
\begin{equation}
J_\m k=2J(\cos k_x+\cos k_y),\;\m{D_k}=2i d(-\sin k_y,\sin k_x).
\end{equation}

	 \begin{figure}
		 \centering
			 \includegraphics[width=14cm]{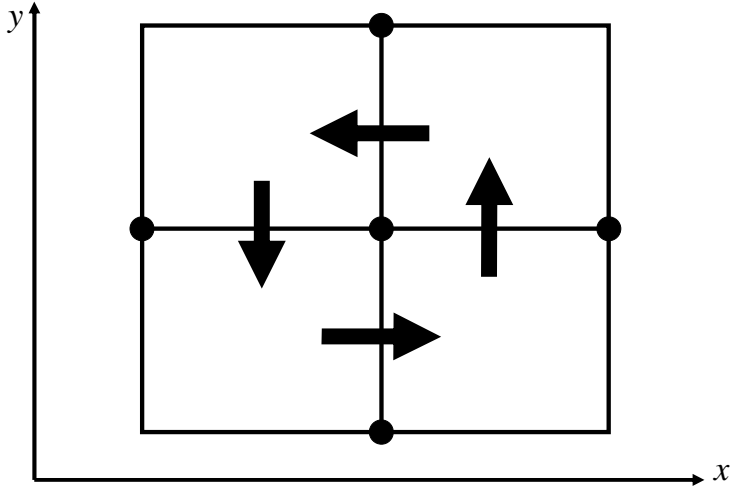}
		 \caption{ Nearest  neighbors and DM vectors in the square lattice.}
		 \label{Fig.3}
	 \end{figure}

\subsection{AF layer}

Using Eqs.(9) and (32) from the conditions $\partial E_a/\partial k_{x,y}=0$  we obtain
\begin{equation}
\tan k_x =-k_0\h c_y,\;\tan k_y=k_0\h c_x,\; k_0=d/J,
\end{equation}
where $\h c$ is given by Eq.(12). At $k_0\ll 1$ we have $ (\h c\cdot\m k)=0$. So  there is the  AF cycloid as in Sec.III.

From Eq.(33) follows
\begin{equation}
\cos k_{x,y}=1/R_{y,x},\;\sin k_{x,y}=\mp k_0\h c_{y,x}/R_{y,x},
\end{equation}
 where $R_{x,y}=\sqrt{1+\h c^2_{x,y}k^2_0}$  and
   we have
\begin{equation}
E_a=-S^2J(R_x+R_y)-S(\h c\cdot\m H)^2/2H_a,
\end{equation}
where $H_a=2S J_0,\;J_0=4J$  [See Eq.(11)].

 At $k_0 \ll 1$  in the in-plane field we have
\begin{align}
E_a=-\frac{S^2J_0}{2}\left\{1+\frac{k^2}{4}-\frac{k^4}{16}[\sin^4\phi+\cos^4\phi]\right\}-\frac{H^2\cos^2(\phi-\Psi)}{2H_a},
\end{align}
where we replaced $k_0\to k$. The first two terms are the AF energy and DMI contribution  respectively. The third tern is the DMI induced square anisotropy.

In zero field $E_a$ is minimal  at $\phi=\pm\pi/4$  and we have two chiral domains with $\h c_\pm=(1,\pm 1)/\sqrt{2}$ and $\m k_\pm=k(\mp 1,1)/\sqrt{2}$ \cite{A}.

The in-plane  field
rotates the $(\h c\cdot\m k)$ domains. As  the  DM anisotropy is of order of $k^4$ there are two field regions. In the strong field when $H\gg H_{a 1}=H_a k^2/4$, the anisotropy may be neglected, the chirality $\h c$ is along the field ($\phi\simeq\Psi$) and the spin flip occurs  at $H=H_a$.

In the weak field ($H\sim H_{a 1}$) the magnetic structure is determined by two last   terms in Eq.(36). In the dimensionless units we have
\begin{align}
F(\phi)&=\sin^4\phi+\cos^4\phi-2W_a\cos^2(\phi-\Psi),\\
\frac{d F}{d\phi}&=2[ -\sin 2\phi\cos2\phi+ W_a\sin(2\phi-2\Psi)]=0,
\end{align}
 where $W_a=(H/H_{a 1})^2$. The rotation of the chiral domains is describes by Eq.(38). We consider   two simplest cases.

  i. $\Psi=0,(\m H||\h x)$. Eq.(38) has two solutions: $\cos 2\phi=W_a$ and $\sin 2\phi=0$. From the first solution we obtain  $\sin\phi=\pm\sqrt{(1-W_a)/2}$ and $F=1/2-W_a-W^2_a/2$. The field rotates the $\h c_\pm$ domains to  $\h x$ axis.  At $W_a>1$ we have  one domain with  $\phi=0$ and  $F=1-2W_a$.  At $W_a=1$ we have the second order  transition to the one  domain state.
	
	ii. $\Psi=\pi/4,(\m H|| \h x+\h y$).  There are two solutions again: $\cos 2\phi=0$ and $\sin 2\phi=-W_a$.
In the first case we have $\phi=\pm\pi/4$ and two    domains with $F(-\pi/4)=1/2$ and $F(\pi/4)=1/2-2W_a$.
At $H>0$ the $-\pi/4$ domain is unstable and we have one domain with $\h c ||\m H$. The second solution must be ignored as $F=[1+(1-W_a)^2]/2\geq 1/2$.
 	
  In the out-of-plane	  field we can neglect the DMI anisotropy.  As a result instead of
Eqs.(20) and (21) we obtain
\begin{align}
E_{a 1}&=-(S^2 J_0 k^2/8)[\sin^2\theta+G\cos^2(\theta-\Theta)],\\
\m k&=k(-\sin\Psi,\cos\Psi)
\sin\theta,
\end{align}
where $G=(H/H_{s f})^2$,   the spin-flop field $H_{s f}=H_a k/2\sqrt{2}\gg H_{a 1}$ and $\phi=\Psi$.

   In  the perpendicular field ($\Theta=0$) at $H=H{s f}$ we have the spin flop    to the AF state as in Sec.III.

		If $\Theta\neq 0$  from  Eqs.(A9,10)   we have
		\begin{align}
		E_{a 1}&=-(S^2 J_0 k^2/16)[1+G+(1-2G\cos 2\Theta+G^2)^{1/2}],\n\\
\sin^2\theta&=\frac{1}{2}\left[1+\frac{1-G\cos 2\Theta}{(1-2G\cos 2\Theta+G^2)^{1/2}}	\right].	
\end{align}
As a result the  length of the cycloid wave vector $|\m k|=k\sin\theta$ depends on the field.  For example  $|\m k|=k/\sqrt{2}$  and $k\sin\Theta$ for $H\cos\Theta=1$ and $G\gg 1$  respectively \cite{A}.	In the last cases as in Sec.III  we have the conical cycloid with $\h c ||\m H$ and the spin flip at $H=H_a$.

\subsection{FM layer}

From Eqs.(10), (11) and (32) we obtain
\begin{align}
\tan k_x&=k_0\h c_y,\;\tan k_y=-k_0\h c_x,\; k_0=d/J, \\
\sin\alpha&=-\frac{(\m H\cdot\h c)}{H_f\sin^2\theta},\; H_f= S J_0 k^2/4.
\end{align}
where  expressions for $\tan k_{x,y}$ have other signs than in Eq.(33),  $\h c$ and $\m H$ are given in Eq.(12).

 In the in-plane  field we have
\begin{equation}
E_f=-\frac{S^2 J_0}{2}\left[1+\frac{k^2}{4}-\frac{k^4}{16}(\sin^4\phi+\cos^4\phi)\right]-\frac{H^2\cos^2(\phi-\Psi)}{2H_f}.
\end{equation}
  This equation coincides with Eq.(36)   after replacement $H_f\to H_a$. So all results obtained above  in the in-plane field are valid  after   replacing $H_{a 1}\to H_{f 1}=H_f k/\sqrt{2}\sim k^3$ and $W_a\to W_f=(H/H_{f 1})^2$.

In the perpendicular field instead of Eq.(26) we have
\begin{equation}
E_{f 1}=-(S H_f/2)[\sin^2\theta +(H^2/H^2_f\sin^2\theta)\cos^2\theta].
\end{equation}
As below Eq.(26) one can show that at $H=H_f/2$ there is the spin flop to the FM state.

 \section{Triangular lattice}

\begin{figure}
	\centering
		\includegraphics[width=14cm]{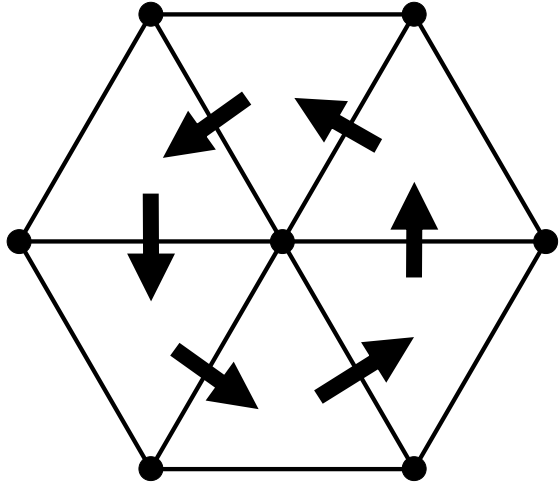}
	\caption{Nearest neighbors and DM vectors in the triangular lattice.}
	\label{Fig.4}
\end{figure}

The nearest neighbor bonds and DM vectors   are shown in FIG.4. From Eqs.(6,9) and  (11)   we obtain
\begin{equation}
E_a=-S^2\sum_{n=0,\pm 1}
\{J\cos(\m b_n\cdot\m k)-
d[(\h c\cdot[\h z\times\m b_n])\sin(\m b_n\cdot\m k)]\}-S(\m H\cdot\h c)^2/2H_a,
\end{equation}
where  $\m b_0=\h x$,$\m b_\pm=\h x\cos\psi\pm\h y\sin\psi$ and
$\psi=\pi/3$.

 As  in Sec.III we have $\m k=k(\cos\xi, \sin \xi)$ and     obtain
\begin{equation}
E_{a 1}=-S^2J\sum(\cos k\cos\xi_n-k_0\sin\phi_n\sin k\cos\xi_n),
\end{equation}
where $k_0=d/J$, $\xi_n=\xi+n\psi$ and $\phi_n=\phi+n\psi$.

In the $k^2$ approximation  we have
\begin{equation}
E_a=-(S^2J_0/2)[1-k^2/4-(k k_0/2)\sin(\phi-\xi)],
\end{equation}
where $J_0=6J$.This energy is minimal at $\xi=\phi+\pi/2$  as in Eq.(14) and we obtain
\begin{align}
E_a&=-(S^2 J_0/2)(1+k^2/4)-S H^2\cos^2(\phi-\Psi)/2H_a,\n\\
\m k&=k_0(-\sin\phi,\cos\phi),
\end{align}
where $H_a=2S J_0$. This expression coincides with  the energy (36) of the square lattice if one  neglects the $k^4$ terms.The same takes place in the FM case where $H_f=S J_0 k^2/4$ [See  Eq.(44)]. So  all results obtained in Sec.V  for the out-of-plane field remain valid as the DM anisotropy may be neglected.

The DMI  hexagonal  anisotropy is  of order of $k^6$.The $k^4$ and $k^6$ terms  of the energy  (47) are studied in Appendix C. In the $k^4$ approximation  we have $E_4=-3S^2 J_0 k^4/128$.  The DM anisotropy is following
\begin{equation}
E_{ A6}=\frac{S^2 J_0 k^6}{ 96^2}(\cos 6\phi-1).
\end{equation}
This energy is minimal at  $\phi=\pm\pi/6$ and $\phi=\pi/2$. We have  three chiral domains with $\h c$ along these directions. The domain rotation field is very weak.  We have $H_{a 1}\sim H_a k^3/96$ and $H_{f 1}\sim SJ_0 k^4/96$. So we do not study their field rotation.

\section{The layer electric polarization}

We consider a possibility of the layer electric polarization in the cycloidal state similar to the observed  in multiferroics (\cite{MU} and references therein). We use the same method as in \cite{SD}.

The layer  is  at a distance $z_0$ above the substrate. It is fixed by an effective potential well $V(z-z_0)$.
The DMI depends  on the layer position  and we have  $d=d(z)$. In the cycloidal stat the total layer energy is following
 \begin{equation}
E_L(z)=V(z-z_0)-S^2J_0 Q k^2(z)/2,
\end{equation}
where $k(z)=d(z)/J$, $Q=1/3$ in the rectangular  lattice and $Q=1/4$  in two other  lattices (See Sec.IV).
The minimum of this energy determines the layer shifting $\delta z=z-z_0$. If  $V(\delta z)=K\delta z^2/2$  we obtain
\begin{equation}
\delta z=(S^2J_0 Q/K) k(z_0)d k/d z_0.
\end{equation}
Due to this shifting the electric polarization $\m P\sim\h z \delta z$
 may appear. It disappears   with the cycloid. One can mention temperature [$S^2(T)\to 0$]
 and the spin flop in the  perpendicular magnetic field. In general  the $\m P$ field behavior is    determined by the factors  $(1+\sin\phi)\sin^2\theta$
 and $\sin^2\theta$ in the rectangular lattice and in two other cases respectively.

\section{Discussion}

We used above  the classical approximation.  Any fluctuations were ignored. Meanwhile in the $2D$ magnets
 they are  very important  and  may destroy the   magnetic order at $T>0$. So the study the spin waves  with the small momenta is the  urgent problem.

In this paper we considered a monolayer  as a mirror breaking surface giving rise the DMI.  In the surface films and interfaces with few layers the mirror symmetry is broken on both sides. As a result the different DMI must be in two boundary layers.       For example in  the interface with the same  material on both sides
 the DM vectors in two boundary layers have opposite directions. The films with two, three and four layers  must have different magnetic structures.  In general the magnetic structure  of the  thin film  depends on the number of the layers. It was observed recently \cite{V}.

\appendix
\section{}

Minimum conditions for Eqs.(16) and (39) coincide after replacement $(\phi,\Psi)\to (\theta,\Theta)$. We consider the second.  In the dimensionless units we have
\begin{align}
E&=-\sin^2\theta-G\cos^2(\theta-\Theta),\\
d E/d\theta&=-\sin 2\theta+G\sin 2(\theta-\Theta)=0,\\
d^2 E/d\theta^2&=2[-\cos 2 \theta+G\cos 2(\theta-\Theta)]>0.
\end{align}
 Eqs.(A2) and (A3)  may be represented as follows
\begin{align}
    -G\sin 2\Theta\cos 2\theta-(1-G\cos 2\Theta)\sin 2\theta&=0,\\
 -(1-G\cos 2\Theta)\cos 2\theta+G\sin 2\Theta\sin 2\theta>0.
\end{align}
From  Eqs.(A4) and (A5) we obtain
\begin{equation}
\sin 2\Theta/\sin 2\theta
>0 .
\end{equation}Solution of Eq.(A4) is following
\begin{align}
\sin^2 2\theta&=\frac{G^2\sin^2 2\Theta}{D},\;\cos^2 2\theta=\frac{(1-G\cos 2\Theta)^2}{D},\\
D&=1-2G\cos 2\Theta+G^2.
\end{align}
 If $\cos 2\theta=-(1-G\cos 2\Theta)/D^{1/2}$ we have
\begin{align}
\sin^2\theta&=\frac{1}{2}\left[1+\frac{1-G\cos 2\Theta} {(1-2G\cos 2\Theta+G^2)^{1/2}}\right],\\
E&=-(1+G+\sqrt{1-2G\cos 2\Theta+G^2})/2.
\end{align}
At $G=0$ and $G\gg1$ we obtain $\sin^2\theta=1$  and $\theta\simeq\Theta$ respectively.

\section{}

 From Eq. (20)
  we have
\begin{equation}
E=-(1+\sin^2\phi)\sin^2\theta-g(\h H\cdot\h c)^2,\end{equation}
where $E=12E_{a 1}/S^2 J_0k^2$, $g=(H/H_{a c})^2$ and $(\m H,\h c)$ are defined in Eq.(12).  The replacement $(\phi,\theta)\to(\phi+\pi,-\theta)$ does not change  Eq.(B1).

  The minimum conditions are following
\begin{align}
 dE/d\theta&=-(1+\sin^2\phi)\sin 2\theta-2g(\h H\cdot\h c)[\cos(\phi-\Psi)\cos\theta\sin\Theta-\sin\theta\cos\Theta]=0,\\
dE/d\phi&=-\sin 2\phi\sin^2\theta +2g(\h H\cdot\h c)\sin(\phi-\Psi)\sin\theta\sin\Theta=0,
\end{align}
 At $g\gg 1$ the $g$ terms must be of order of unity. As a result we have   $\phi\simeq\Psi$ and $\theta\simeq\Theta$.

\section{}
We evaluate below the DMI anisotropy in the triangular lattice.
 From Eq.(47) in the  $k^4$ approximation  we obtain
\begin{equation}
 E_2+E_4=-\frac{S^2J_0}{2}\left[1-\frac{k^2}{4}-\frac{k k_0\sin(\phi-\xi)}{2}-\sum\left(\frac
{k^4\cos^4\xi_n}{4! 3}-\frac{k^3 k_0\sin\phi_n\cos^3\xi_n}{3! 3}\right)\right].
\end{equation}
This  energy is minimal at  $\xi=\phi+\pi/2$ asnd  $k=k_0$ [See Eqs.(48) and (49)].  Taking into account that $\sum\sin^4\phi_n=9/8$  we  obtain
\begin{equation}
  E_4=-3S^2J_0/128.
\end{equation}
By the same way we obtain
\begin{equation}
E_ 6=-\frac{S^2 J_0 k^6}{432}\sum\sin^6\phi_n.
\end{equation}
From this expression for the DMI anisotropy
we obtain
\begin{equation}
E_{A 6}=\frac{S^2J_0 k^6}{96^2}(\cos 6\phi-1).
\end{equation}


\begin{thebibliography} {99}
\bibitem{CC} More general  $\m b=\pm(\mu\h x\pm\nu\h y),\;\cos^2\rho=\mu^2/(\mu^2+\nu^2)$.If $\mu=\nu$ we have the square  lattice considered in Sec.V.
\bibitem{A} As explained below Eq.(11) we consider the pair $\pm(\m k,\h c)$ as single state.
\end{thebibliography}
\end{document}